# Untangling Cognitive Processes Underlying Knowledge Work

Ginar S. Niwanputri
The University of Sheffield, gsniwanputri1@sheffield.ac.uk

Elaine Toms
The University of Sheffield, e.toms@sheffield.ac.uk

Andrew Simpson
The University of Sheffield, andrew.simpson@sheffield.ac.uk

In a post-industrial society, the workplace is dominated primarily by Knowledge Work, which is achieved mostly through human cognitive processing, such as analysis, comprehension, evaluation, and decision-making. Many of these processes have limited support from technology in the same way that physical tasks have been enabled through a host of tools from hammers to shovels and hydraulic lifts. To develop a suite of cognitive tools, we first need to understand which processes humans use to complete work tasks. In the past century several classifications (e.g., Blooms) of cognitive processes have emerged, and we assessed their viability as the basis for designing tools that support cognitive work. This study re-used an existing data set composed of interviews of environmental scientists about their core work. While the classification uncovered many instances of cognitive process, the results showed that the existing cognitive process classifications do not provide a sufficiently comprehensive deconstruction of the human cognitive processes; the work is quite simply too abstract to be operational.

## 1 Introduction

Today's workplace is dominated by *Knowledge Work* (KW) [13], work that deploys data and information to create new information and knowledge. The outcomes may range from a simply oral response to a formal document or crucial decision. To achieve that outcome, workers must execute and engage multiple cognitive processes (CPs) with support from software applications, e.g., word processors, dictionary, spreadsheet, calculator. While tools deal with complex operations such as statistical analysis and searching for documents in large databases, many CPs that require the manipulation of data and information remain unsupported. Consider the typical tasks of assessing a student essay, evaluating a research proposal, or comprehending a complex medical or legal outcome. Each will use many CPs that engage with data and information on many levels. The challenge is what are those CPs and how do they work together?

A CP is any …" mental function assumed to be involved in the acquisition, storage, interpretation, manipulation, transformation, and use of knowledge" [APA https://dictionary.apa.org/cognitive-process]. They are evident in most information-dominant activities such as report writing, planning and decision making, although few descriptions of work tasks provide that level of detail. To date, most of the work that has examined CPs in work have occurred in education such as the work of Bloom and colleagues [1, 2, 10, 20]. We speculated whether these classifications could form the foundation in understanding CPs at a sufficiently detail level to identify the requirements for a suite of tools.

## 2 RELATED WORKS

### 2.1 KW and Task

KW is an often-maligned concept as it is often ascribed only to those who manage, organise, and disseminate information. Its use in this research pertains to those whose tasks manipulate data and information using CPs as the key component in achieving work goals [3, 6, 20]. In essence, the workers primarily 'think' for a living. Although cognition is present in all forms of human activity, it is arguably the mental part that plays a more significant role in KW, than in other types of human activity [19]. Work has become more challenging with digitalisation and people are increasingly required to handle what now looks like unfamiliar tasks [16].

Although KW is mainly conceptual, it still is composed of discrete sets of tasks which are those actions a user takes to achieve a goal. Most of these tasks such as write a dissertation proposal are broken down into subtasks, that include designing a method and developing a review of past work. Each of these subtasks uses many CPs [19] that manipulate information [9]. Most prior work have focused on work intensification and automation, with limited empirical research on cognitive aspects [9,17].

## 2.2 CPs

To understand the CPs that underpin KW, we need to understand human cognition which can be defined as all processes by which the sensory input is transformed, reduced, stored, recovered, and used [17]. Human cognition is a collectively specialised set of mental operations that can process information from the environment and results in adaptive behaviours across various contexts [14].

Several different models have been used to investigate human cognition. Broadbent [4] introduced a model outlining several human brain information processing phases [11]. Technology has profoundly influenced how human cognition is defined in transforming the mind into an information-processing model or device [12,14]. Some scholars see it as a model for processing information [4,12,15], complete with inputs, various processing methods, and outputs. This input/output interpretation is a metaphor for the brain's functioning that is widely used but not entirely supported by scientific research (for example, [8]).

These models present cognition at an abstract level but they do not consider how these models may be deconstructed and operationalised. Bloom et al. [2] developed a hierarchy of learning levels based on CPs (knowledge, comprehension, application, analysis, synthesis, evaluation). Anderson & Krathwohl [1] modified Bloom's version which resulted in a variation that included six differently labelled levels: Remember, Understand, Apply, Analyse, Evaluate, and Create. This version added knowledge dimensions for each level giving a more comprehensive view than Bloom's. A third classification is that of Cattell-Horn-Carroll (CHC) [7] which consists of limited abilities, broad abilities, and general abilities in cognition [10, 18]. They represented CP as any action consisting of mental content executed to deliver a response as a result, focusing on cognitive abilities. In a fourth version, Wang et al. [21] expanded the human cognition model into a Layered Reference Model of the Brain (LRMB). In LRMB, natural intelligence's physiological mechanisms and CP are explained formally and rigorously. All layers may be broken down into its 39 components, organised into six tiers: sensation, memory, perception, action, meta-cognition, and higher cognition. The two conscious layers, Higher CPs (HCP) and Meta CPs (MCP), interact with conscious life functions. Other layers are subconscious and can be grouped into Sensational CP (SCP) and Subconscious CP (SCCP), consisting of activities such as memory, perception, and action. Higher CP depends on activities such as perception, memory, abstraction, and the pace at which the interdependent processes are used.

By reviewing prior CP classifications, Bloom provided a more definitive list of processes. However, these processes are conceptual and broad and consequently do not provide much guidance regarding how instruments should be built to promote those processes. CHC mainly focused on mapping various cognitive abilities that compose individual intelligence, which was utilised to develop the baseline of cognitive measurement, but not to identify CP. LRMB provided a vast number of CP. Their classification is still considered inadequate for comprehensively uncovering CPs behind KW since only the two conscious layers help provide some parts of CPs. In sum, these models are excessively abstract, with little information that can be applied to explain how KW is carried out or could be assisted. Our knowledge of them is too broad to allow us to unbundle the detail of mental processes under work tasks.

## 3 RESEARCH OBJECTIVES

While cognitive aspects of work are evident in any workplace, surprisingly, limited research has focused on uncovering the CPs used in KW. As discussed earlier, we have a surplus of classifications at varying levels of specificity, but they appear to be insufficient to serve as the foundation for how we design and develop specific tools to assist predominantly cognitive tasks. This research puts these classifications to the test by examining in a small data set what CPs are used in KW, and what are the merits of these classifications in describing KW CPs.

This research was guided by two research questions: (1) What sorts of CPs are typically used in KW? and (2) How effective are existing classifications in deconstructing those processes?

## 4 METHODS

To understand the nature of KW, we needed rich descriptions of tasks. Rather than collect another data set, we reused a previously unpublished secondary data set that contained semi-structured interviews conducted with environmental scientists who work in coastal and ocean fields and work for federal/provincial governments, non-governmental organisations, academic institutions, and consulting firms. They were recruited through professional networks and their work roles ranged from those with less than two years of working experience to managers and directors with more than ten years of experience. The interviews deployed the critical incident technique in which participants were asked to recall and describe a significant project they had previously worked on. The interviews lasted between 29 and 71 minutes with extra follow-up for specific details and clarification.

There are two issues of concern with respect to the data. First, these interviews focused on work tasks and how information was acquired and used in the task but did not specifically consider CPs. Second, the data were collected more than ten years ago.

Because we were looking at a 'proof of concept' we accepted that the interviews may not have dealt with the matter at a sufficient depth. Because we were more concerned with how people achieve a task, our analysis focused on the tasks which were timeless and not the tools they might have used; we were unconcerned about the age. The environmental problems remain the same, e.g., is this a good location for a wharf?

The data analysis process took the following steps. First the key aspects of the interview that dealt specifically with the work task were extracted from the transcribed interviews; this resulted in 195 segments across 17 tasks. Second, each of the segments was analysed using the procedure documented in Table 1. Third, each of these segments was mapped to one or more CPs which were extracted from the LRMB [21] and A&K's [1]. See Table 2 for an example. The complete list of 34 possible CPs is illustrated in Table 3, along with the occurrences of each.

**Table 1** How each segment was analysed (Example)

| Step | Action | Explanation |
|---|---|---|
| 1 | *Select* a segment. | "So I would call the area staff, and I would call the lobster biologist …… and I talked to my own staff who had some understanding of the situation." |
| 2 | *Analyse* the segment by considering the meaning, the action and the context and then infer a suitable CP. | Define the right persons to call/talk can be inferred into H-Decision making |
| 3 | *Map* the segment to a CP code | Decision-making is choosing a course of action from a set of alternatives. Here the "course of action" represents the right persons, and "the set of alternatives" are: {list of persons}. Thus, the segment has been mapped to the definition of the CP decision-making and has been encoded. |

**Table 2** Samples from the coding scheme

| CP | Definition | Level |
|---|---|---|
| Analysis | Higher-CP of the brain divides a physical or abstract object into parts to examine or determine their relationship deductively. | HCP |
| Searching | Meta-CP of the brain based on trial-and-error explorations to find a set of correlated objects, attributes, or relations for a given object or concept; or to find valuable solutions for a given problem. | MCP |
| Action | A set of subconscious CPs of the brain executes bodily (external) and mental (internal) actions via the body's motor systems or the brain's perception engine. | SCCP |
| Vision | The sensational CP of the brain detects and receives visual information from entities of the external world in forms of images, shapes, sizes, colours, and other attributes or characteristics. | SCP |

Much of the analysis was *inferential.* Participants did not specifically use the words specified in list of processes as this is not the typical way that one describes one's work. For example, they are unlikely to say, "I made a decision to call my colleague" and more likely to say "I contacted a colleague." Codes were assigned by interpretation of intent.

## 5 FINDINGS AND ANALYSIS

The findings are presented in two parts. The first illustrates the types and frequencies of CPs that were mapped to the segments, while the section illustrates how these processes map back to a task.

*Part 1. Occurrence of CPs*

The results from classifying the segments according to four groups: Higher, Meta, Subconscious and Sensational CP are illustrated in Table 3 which identified 469 instances of a CP used across all 195 segments from 17 tasks. The Higher CPs (in the left column) show that comprehension and decision making accounted for 83 or 47% of all Higher CPs. At the Meta stage (identified by #2 in the table), shows that Searching was the most frequently used, while at the Subconscious, (i.e., #3) it was all about Action, and at the Sensational (#4), audio and visual.

Tables 3 illustrates the number of occurrences of CPs. In general, 34 CP types occurred at all levels across the 17 work tasks. Most of the cognitive effort occurred in Higher-CP: comprehension, decision making, analysis and evaluation. At the Metacognitive level, 97 instances were observed; most of this effort (i.e., 68 or 70%) was devoted to searching and memorisation. These would have been activated by one of processes identified in either Higher CP or Subconscious CP. Action which accounted for 77 instances of Subconscious CP is the most numerous CP, likely because it is a part of any task or subtask. Action will always accompany other CPs in Higher or Metacognitive layers. Searching is the second-most frequently observed CP, largely because most work tasks involve research. As a subtask type, information-gathering responsibilities may occasionally be included in work activities [5].

Table 3 Frequency of occurrence of CPs by type

| 1. Higher | | 2. Meta | |
|---|---|---|---|
| Comprehension | 47 | Searching | 59 |
| Decision Making | 36 | Memorisation | 9 |
| Analysis | 29 | Categorisation | 8 |
| Evaluation | 27 | Abstraction | 7 |
| Creation | 19 | Attention | 7 |
| Explanation | 19 | Knowledge Representation | 4 |
| Planning | 15 | Concept establishment | 3 |
| Imagery | 14 | **3. Subconscious** | |
| Recognition | 11 | Action | 77 |
| Induction | 8 | Motivation | 5 |
| Analogy | 7 | Emotions | 3 |
| Application | 6 | Self-consciousness | 3 |
| Quantification | 5 | Goal setting | 1 |
| Synthesis | 5 | Memory | 1 |
| Learning | 3 | Willingness | 1 |
| Deduction | 2 | **4. Sensational** | |
| Problem-solving | 2 | Vision | 14 |
| Reasoning | 1 | Audition | 11 |

These findings support the view that KW is a bundle of cognitively intensive information processes. That this set of tasks was diverse in its use of information may not come as a surprise. Analysis, evaluation, and comprehension were captured mostly before decision-making. This was followed by the creation which appeared mostly under document development.

*Part 2. Mapping CP to Task*

As another view to examining CPs, we mapped the processes back to the specific task as illustrated in Figures 1 and 2. Each illustration outlines the task goal along with the specific subtasks. In both examples, the CPs are illustrated for only the first subtask. This result challenges our initial hypothesis of using an existing set of CPs to understand the inner workings of a task. While many CPs were identified, it remains unclear as to how these provide insights into the cognitive processing that occurs in a task. These limitations show the inadequacy of using these CPs to assist with identifying the requirements of a tool that might facilitate a particular subtask.

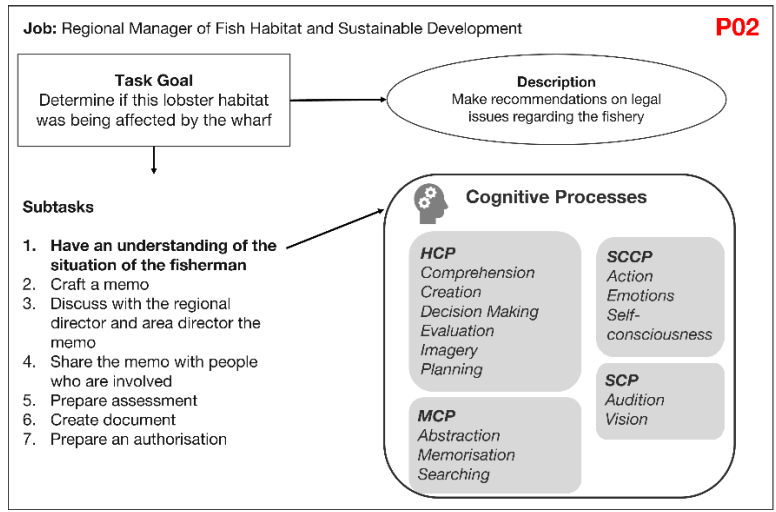

**Figure 1 CP in Participant 02 - Have an understanding of the situation of the fisherman**

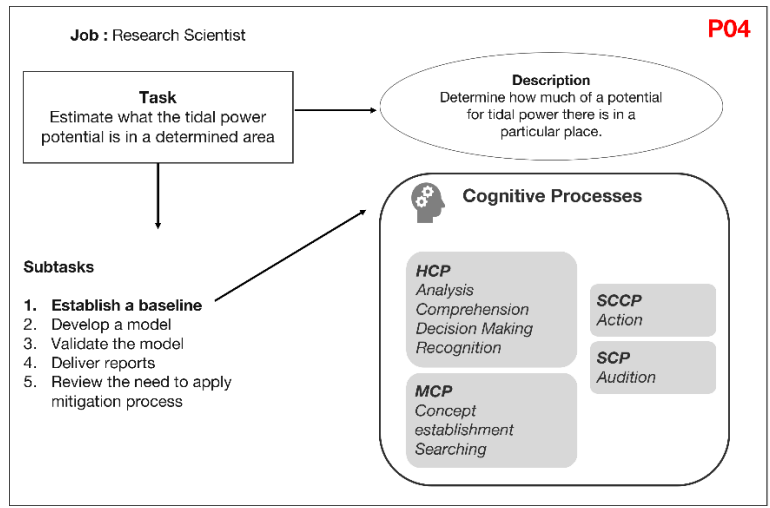

**Figure 2 CP in Participant 04 - Establish a baseline**

# 6 CONCLUSION

This research identified a set of CPs that were used in a set of tasks completed by environmental scientists. Clearly processes like comprehension, decision making and searching are core to the work. While the task descriptions may have lacked in detail the essence of the task, the application of existing classifications of CPs demonstrated the limitations of those classifications. They are simply too abstract (more of an intellectual exercise) and thus too limited to operationalise in any form of software engineering.

To provide more tools that either automatic, semi-automate or simply support the user in task completion remains an under researched problem. But we first need a more comprehensive understanding of what a person actually – cognitively and physically -- does when doing KW. Our long-term goal is to develop a suite of cognitive tools that will augment human cognitive capabilities, enhance cognitive ability or reduce cognitive effort. KW needs cognitive tools that augment and enrich information processing for humans in much the same way that search engines have done for information findability. This work is of some urgency before newer technologies tells the user what to do and/or controls what and how the user does the work.

## ACKNOWLEDGMENTS

This research is funded by the DoSSIER project under European Union's Horizon 2020 research and innovation programme, Marie Skłodowska-Curie grant agreement No 860721.